# Simulating Forces

## Learning Through Touch, Virtual Laboratories

Felix G. Hamza-Lup, Faith-Anne L. Kocadag
Department of Computer Science and Information Technology
Armstrong Atlantic State University
Savannah, Georgia
e-mail: felix.hamza-lup@armstrong.edu, fk4687@stu.armstrong.edu

*Abstract*— With the expansion of e-learning course curricula and the affordability of haptic devices, at-home virtual laboratories are emerging as an increasingly viable option for e-learners. We outline three novel haptic simulations for the introductory physics concepts of friction, the Coriolis Effect, and Precession. These simulations provide force feedback through one or more Novint Falcon devices, allowing students to "feel" the forces at work in a controlled learning environment. This multimodal approach to education (beyond the audiovisual) may lead to increased interest and immersion for e-learners and appeal to the kinesthetic learners who may struggle in a traditional e-learning course setting.

*Keywords-Haptic simulations; e-Learning*

## I. INTRODUCTION

E-Learning has exploded in popularity in recent years, and for good reason. Both online and brick-and-mortar institutions offer an increasing variety of courses on the web to students from around the world. While the convenience of an e-learning course is difficult to beat, instructors may struggle to retain students, keep them engaged, or know whether their students are fully grasping the material. Furthermore, many courses do not always translate effectively into existing Edtech (Education Technology) platforms [1, 2].

As virtual classrooms proliferate, the tools of trade continue to develop in tandem. Multimodal interactions are especially important, and these novel and multidimensional approaches have been proven to increase user engagement, interaction, and mastery of concept [3–6]. While these virtual classrooms do not replace traditional face-to-face teaching models, they can augment these models and may prove invaluable to e-learning course curricula.

Haptics in computing refers to the addition of force feedback to the user through commercially available hardware. Through this technology, users may engage their senses beyond their visual perceptions alone, allowing for a more intuitive understanding of complex or abstract concepts. Haptics in virtual laboratories are particularly effective when touch is required for the correct comprehension of physical phenomena, variation of frequencies, medical procedures, engineering, virtual museums, etc. [4].

We provide an overview of three haptic-based virtual simulators that can be merged into existing Edtech systems like Vista or MOODLE. These simulators take advantage of the open source H3D API, creating three dimensional audiovisuals coupled with a tactile (haptic), interface. The three simulations outlined in Section 2 augment the teaching of Introductory Physics Concepts of: friction, the Coriolis Effect, and torque-induced precession [7].

## II. BACKGROUND AND RELATED WORK

### A. E-Learning and Virtual Laboratories

While the majority of e-learning programs are merely video, chat, and discussion board based, it is easy to see the prudence of elevating to a standard that may nurture and stimulate students' curiosity and aptitudes. Creating an authentic learning experience has long been a concern of e-learning course providers, and many experts agree that such an environment requires community, "experimentation and action" [8].

Haptic, or kinesthetic, learners are those who prefer a more active approach to course materials [9]. Vincent and Ross estimate that these kinesthetic learners make up approximately 17% of the population [10]. The integration of virtual laboratories into online Edtech platforms creates environments where e-learners may both self-teach and collaborate with others to maximize their learning potential [5].

Brown, et al [11], argue that how a person perceives an activity is dependent on their environments and tools. Thus the implementation of haptics in e-learning may improve experiments where the representation of material properties and experimentally relevant forces are of the utmost importance [4]. Dudulean et al found that haptic feedback, through a low cost and relatively small device, increased the effectiveness of an interaction, resulting in students spending more time exploring the virtual objects, and increased motivation, interest, critical thinking development, and problem solving [12]. While most haptic simulations were designed to augment traditional classrooms, Schaf, et al [6] went one step further to integrate their deriveSERVER (providing remote access to a virtual reality environment and also a real experiment) with a collaborative MOODLE interface for their engineering workspaces.

The ideal collaborative learning environment for engineering education, according to Pereira et al [5], includes: a shared workspace for educational media and a theoretical material module (common to virtual learning environments), an immersive 3D social interface (like SecondLife), content adaptation to user feedback, integration of virtual labs or experiments, intelligent tutoring systems, teamwork and collaboration support, augmented sense immersion (beyond just sight, hearing, and touch), and serious game concepts – the use of game-like solutions that capture attention and educate as they entertain [5]. While no such system yet exists, the continued incorporation of haptic technology into existing e-learning courseware may be a great step toward providing distance learners an education more on par to that of students at traditional brick-and-mortar institutions.

*B. Haptics APIs: H3D*

SenseGraphics' H3D API is an open source, cross platform development toolkit for creating visuo-haptic scene graphs [13]. It is released under the GNU GPL license with commercial licensing options. The high level interfaces of the API are X3D (another open source format) and Python. While X3D provides the 3D graphics vocabulary, Python describes the application's user interface behavior [14]. Most importantly, H3D allows for rapid prototyping and supports a wide range of haptic devices.

III. CASE STUDIES

The three physics demos outlined herein were developed with the intention of augmenting the introductory (calculus-based) physics curricula at Armstrong Atlantic State University in Savannah, Georgia. While the simulations have not yet been implemented into an online e-learning system, expansion into that realm would be an immediate future extension.

Each of the case studies below employed one or more Novint Falcon devices. This device, classified as a game controller, was chosen because of its robustness, relatively small working volume, commercial availability, and increasing affordability. Currently, one can purchase one such device (with the standard features) on the Novint website for the same price as a HD web cam [15].

*A. Concept: Friction*

The Friction demo, detailed in [7], provides a carefully controlled environment where students can perceive the effects of static friction, kinetic friction, slope inclination (and gravity by extension), mass, and user-generated forces on the movement of a block on an inclined plane. While the virtual environment is 3D, the block movement on the inclined plane is restricted to one dimension to facilitate user control. The three dimensionality of the simulation ultimately comes into play through manipulation of the rotating disk at the bottom, center (Figure 1) that allows the user to rotate the scene to view it at different angles.

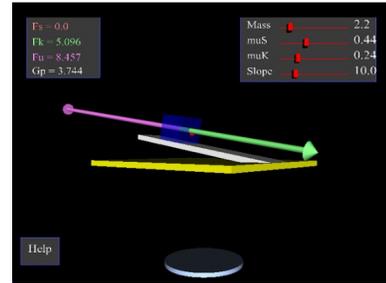

Figure 1. Friction Simulation: Screen shot of simulation.

Students were given instructions on how to interact with a block on an inclined plane. A static frictional force acted on the block to impede its movement, while a kinetic frictional force acted on the block as it moved. Users attempted to move the block via the haptic pointer. In addition to the haptic force feedback from the Novint Falcon hardware, resulting force directions and magnitudes were displayed visually through three dimensional arrows while a heads-up display stated the explicit magnitude values, as illustrated in Figure 2.

An evaluative pre-test of the 86 participants showed that most students had only a rudimentary knowledge of static and kinetic friction, with the average score being 36.7% (random chance would yield a score of 19.7%). After the pre-test, the students attended a 50-minute conventional lecture about static and kinetic friction. The lecture was followed by a post-test, and students were split into two groups with equivalent post-test results.

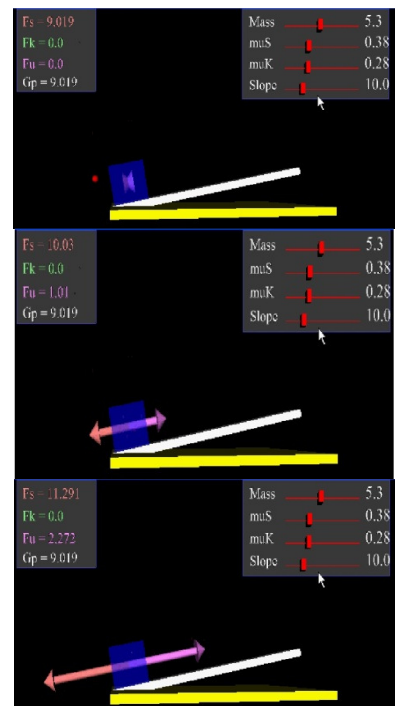

Figure 2. Friction Simulation: Force magnitudes represented as arrows, as the user pushes the block up. The dot on the topmost image near the block represents the position of the haptic pointer.

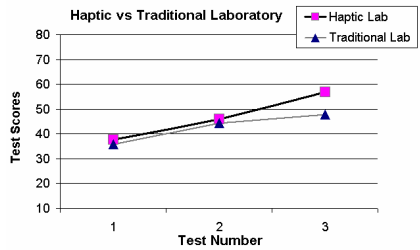

Figure 3. Average Test Scores For Haptic vs. Traditional Lab Groups [7]

After the division into groups, group A performed lab experiments using the visuo-haptic simulator while students in group B performed similar experiments in a traditional laboratory setup. Afterward a final test was administered, test score normalized gains were calculated as

$$(\text{Test 3} - \text{Test 2}) / (100 - \text{Test 2}). \quad (1)$$

Figure 3 illustrates the efficacy of the haptic simulation over traditional teaching methods regarding frictional force concepts. The normalized gain of group A was 0.182, while the gain was slightly negative for group B at -0.011. Not only were average test scores higher among the student users of the simulation, but overall student curiosity and attention measured in an attitude survey were superior to those who had not used the setup.

### B. Concept: The Coriolis Effect

The Coriolis effect is one of the more complex concepts to convey to introductory students. It is a phantom force that appears to alter the path of an object in juxtaposition of another spinning frame of reference. A plane flying south from the North Pole would appear to be deflected to the right (or westward) because of the Coriolis effect.

The Coriolis application attempts to illustrate the concepts of this perceived force through a simple simulation where the user attempts to push a ball into a goal (using the Novint Falcon haptic device) within a spinning frame of reference. While the background of the simulation spins, users feel a deflecting force (representing the Coriolis effect) parallel to the direction of rotation. Users are forced to compensate for this force to score a goal (as illustrated in Figure 4).

In contrast, users may appreciate the change in "feel" without the Coriolis effect - second simulation. The second simulation implements a glider (instead of a ball) that is not affected by the surface friction of the ground, thus mimicking a static (non-rotating) frame of reference.

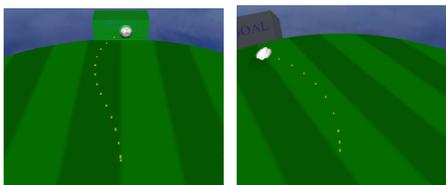

Figure 4. Coriolis Simulation: Ball and Glider Simulations

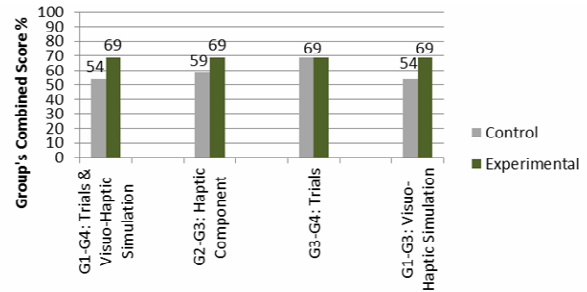

Figure 5. Coriolis Quiz Score Comparisons

24 undergraduate students taking Principles of Physics I at Armstrong Atlantic State University were divided into four groups of six students. GPAs between groups were similar. All groups were given supplemental reading material and a video on the Coriolis effect. Group 2 participated in a visual simulation with no haptic feedback. Group 3 participated in a visuo-haptic simulation involving force feedback. Group 4 was given a tutorial on the use of the haptic devices, then participated in a visuo-haptic simulation with force feedback. All groups were quizzed and given subjective assessment questionnaires at the end.

As shown in Figure 5, the groups that participated in the visuo-haptic simulation showed a 15% advantage in quiz scores over the groups only given reading material and a video. The group that participated in a simulation without haptic feedback only showed a 10% increase in quiz scores. A tutorial on the haptic hardware prior to the simulations did not affect quiz scores, proving either the tutorial ineffective or unnecessary. Both test scores and students' subjective assessments reflected the positive benefits of the simulation, including increased student engagement and grasp of abstract concepts [16].

### C. Concept: Precession

Torque-induced precession refers to the wobble that occurs when a spinning object's axis of rotation shifts in orientation because of an applied torque, or rotational force. Precession is often observed in spinning tops and gyroscopes. Precession, and its relationship to angular velocity and angular momentum, is an important abstract concept that is not always immediately understood, especially by kinesthetic learners. The Gyroscope application provides force feedback through an interactive gyroscope that tilts as it spins (Figure 6,7).

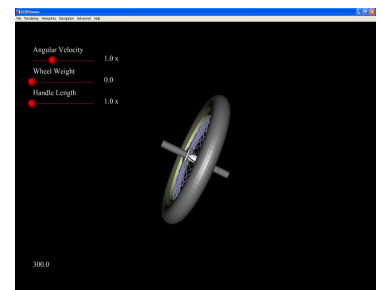

Figure 6. Gyroscope Simulation: Screen Shot

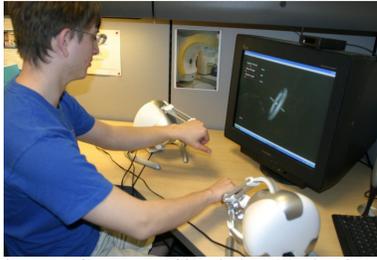

Figure 7. Gyroscope Simulation: Device Setup

This simulation utilizes two Novint Falcon devices, one for each of the user's hands, pointed toward each other. The devices are engaged simultaneously, allowing the user to feel the tilt of the gyroscope handles as the wheel spins. Users can adjust angular velocity, wheel weight, and handle length to experience the resulting precession changes.

The application is currently under assessment and if successful, it may become an integral part of our introductory physics Touchable Virtual Laboratories.

## IV. CONCLUSION AND FUTURE WORK

In the early days of e-learning, costs and lack of sophistication in online courses were prohibitive. The lowering of hardware prices, dramatic improvement of internet bandwidth and reliability, and increased savviness of online educators in their course designs suggest that the popularity of e-learning programs can only grow. While conventional institutions of higher education are not expected to fall by the wayside, they must improve content and knowledge delivery to keep up with the new demands in the informational age. As much as many e-learning instructors are "motivated by a strong conviction that the work they are doing is important to students who need flexible access to education", they are still "clearly meeting a need" [17]. For e-learning courseware to truly compete with the traditional brick-and-mortar programs, measures must be taken by institutions to impart a more immersive, engaging experience on their students.

Just as the pedagogy of physics was advanced dramatically by the introduction of computers as visual learning devices, the tactile activities envisioned in a haptic-enabled laboratory promises similar benefits, especially for kinesthetic learners and for students with disabilities [18]. The three applications outlined in this paper are just examples of the plethora of content a virtual haptic enhanced laboratory can provide. These virtual labs stimulate multiple user senses, and may prove invaluable additions to existing e-learning systems, improving their information distribution capacity, user engagement, and users' learning efficiency.